\documentclass[twocolumn,aps,prd,nofootinbib,tightenlines]{revtex4}

\usepackage{amsmath, amsfonts, amsthm, amssymb, graphicx}

\allowdisplaybreaks[3]

\def\be{\begin{equation}}
\def\ee{\end{equation}}
\def\ba{\begin{eqnarray}}
\def\ea{\end{eqnarray}}
\newcommand{\nn}{\nonumber\\}
\newcommand{\ud}{\mathrm{d}}
\def \pd {\partial}

\begin{document}

\title{Galileons with Gauge Symmetries}

\author{Shuang-Yong Zhou\footnote{ppxsyz@nottingham.ac.uk}}
\author{Edmund J.~Copeland\footnote{ed.copeland@nottingham.ac.uk}}

\affiliation{School of Physics and Astronomy,
University of Nottingham, Nottingham NG7 2RD, UK}

\date{\today}

\begin{abstract}

Galileon models arise in certain braneworld scenarios as modifications to General Relativity, and are also interesting as field theories in their own right. We show how the galileon model can be naturally generalized to include local gauge symmetries, by allowing for couplings to Yang-Mills fields. The resulting theories have at most second order spacetime derivatives in any representation of the gauge group, thereby avoiding Ostrogradski ghosts. We also extend the models to include curved space, and show how in that case we need to include non-minimal couplings between the galileons and the curvature tensors for the theory to retain its second order nature.

\end{abstract}


\maketitle

\section{Introduction}

Recently, there has been plenty of interest in galileon models, either as modifications of gravity or as effective field theories (see~\cite{Trodden:2011xh, Clifton:2011jh} for detailed reviews). Galileons differ from conventional scalar fields in that they possess a Galilean field symmetry
\be \label{galsym}
\pi \to \pi + b_\mu x^\mu + c,\quad b_\mu,~~c={\rm const.}  ,
\ee
hence the name~\cite{Nicolis:2008in}. This field symmetry was motivated by the braneworld Dvali-Gabadadze-Porrati (DGP) model~\cite{Dvali:2000hr}, whose behavior within the Hubble horizon can be captured by a 4D boundary effective field theory where the General Relativity tensor modes are linearized and many distinct features of the model are carried by a galileon that encodes how the DGP brane is bent in the bulk~\cite{Luty:2003vm, Nicolis:2004qq}. Remarkably, this generalized shift symmetry of the field is quite constraining in terms of constructing all possible Lagrangian terms, if we also require there are no higher than second order derivatives in the equation of motion. This second order restriction is important, as we require a well-defined Cauchy problem and higher order derivative terms generally give rise to Ostrogradski ghosts (see, e.g.~\cite{ostroghost}). In four dimensions, there are only 5 possible galileon terms~\cite{Nicolis:2008in}. Unlike the DGP model~\cite{Luty:2003vm, dgpghost}, the general galileon model can give rise to a ghost-free self-accelerating branch~\cite{Nicolis:2008in}. From the perspective of effective field theory, the galileon model has a healthy energy gap between the classical non-linear scale and the quantum scale around a heavy background~\cite{Nicolis:2004qq, Nicolis:2008in}, analogous to that of General Relativity, and the couplings of the galileon interactions are not renormalized by loop corrections~\cite{Nicolis:2004qq, Hinterbichler:2010xn}.

As a surprising and interesting application, the galileon construction has proven crucial in building consistent massive gravity models. For a long time, a consistent massive gravity theory seemed unattainable, due to problems such as the Boulware-Deser (BD) ghost~\cite{Boulware:1973my}. Recently, inspired by the galileon construction, non-linear massive gravity without the BD ghost has been proposed~\cite{arXiv:1006.4367, arXiv:1007.0443, deRham:2010kj, HassanRosen}, where the helicity 0 mode of the massive gravton is exactly a galileon.

Due to these novel properties and appications, interest has spread to consider more general galileon models, motivated by their phenomenologies, 4D symmetries or braneworld scenarios. Since the galileon can be identified as a brane bending mode, multiple galileons may arise in a braneworld scenario with higher co-dimensions~\cite{psz12}. 
General multi-galileon models and their phenomenologies such as self-acceleration and self-tuning have been studied~\cite{psz12}. In~\cite{Deffayet:2010zh}, $p$-form galileon models have been proposed where the galileon scalars are generalized to be differential forms. In certain braneworld scenarios, if a higher co-dimensional bulk has some global symmetry, or isometry, the resulting 4D effective galileon model may inherit this symmetry, leading to a galileon model with global internal symmetry~\cite{Padilla:2010ir, Hinterbichler:2010xn, Zhou:2010di, Andrews:2010km}. Indeed, in this case, the galilean field symmetry and the internal symmetry can both come from the bulk Poincar\'e symmetry. The supersymmetric extension of the galileon model has been studied in~\cite{Khoury:2011da} and an extension to the most general scalar field theory that is of second order in nature has been proposed in~\cite{Deffayet:2011gz, Charmousis:2011bf, horndeski}.

The Galilean field symmetry (\ref{galsym}) may originate from some underlying general covariance. This is indeed the case in the DGP model~\cite{Luty:2003vm, Nicolis:2004qq}. Therefore, the Galilean field symmetry may be ultimately relaxed, and we may consider a generalized symmetry, of which the Galilean field symmetry is some appropriate limit. 
An infrared completion of the galileon model has been considered along the lines of promoting the galilean field symmetry to that of a conformal symmetry~\cite{Nicolis:2008in}, while Deffayet {\it et al} have constructed the covariant galileon by directly covariantising the galileon terms~\cite{Deffayet:2009wt}. The general galileon model constructed by 4D symmetry considerations, as well as its conformal extension and direct covariantisation, has been shown to arise in a probe brane setup where the brane and bulk are endowed with the Lovelock generalisation of General Relativity~\cite{deRham:2010eu}. The general galileon model is the small field limit for a 3-brane in a 5D Minkowski bulk, while the conformal galileon arises for the case of an anti-de Sitter bulk. When the induced brane metric is allowed to fluctuate, the covariant galileon is also recovered. More general brane and bulk settings have been considered, leading to more generalized galileon field theories in 4D~\cite{GHT2011}.

On the other hand, the galileon models contain higher order derivative terms in the Lagrangian, which are stable under quantum corrections~\cite{Nicolis:2004qq, Hinterbichler:2010xn} and may be used to construct solitonic solutions~\cite{Padilla:2010ir, Endlich:2010zj}. In particular, an explicit soliton with a non-linear sigma model constraint has been obtained in the SO(4) symmetric galileon~\cite{Padilla:2010ir}. Further discussions of the gravity and field theoretic aspects of the galileon models may be found in~\cite{Babichev:2010kj, Hui:2010dn, Nicolis:2010se, Pujolas:2011he}. The galileon models have also sparked plenty of interest in cosmology, when considering both the late~\cite{latecos} and early universe~\cite{earlycos}.

As we have mentioned, a global geometric bulk symmetry in certain braneworld scenarios can induce a global internal symmetry on the corresponding 4D effective galileon theory. However, generically, the bulk geometric symmetry may be realized locally, i.e., there is some unbroken general covariance alongside with possible isometries in the bulk. In this case, the resulting effective field theory would inherit a gauged internal symmetry. Indeed, from the formalism of a probe brane plus an induced metric of~\cite{deRham:2010eu}, this is what we would have when there is a bulk metric whose symmetries amount to replacing the partial derivatives in the induced brane metric with gauge covariant derivatives.

In this paper, we consider an important generalisation of the galileon model, asking the question whether the galileons can be consistently coupled to gauge fields whilst maintaining the attractive features such as avoidance of Ostrogradski ghosts. In Section~\ref{galgb}, we briefly review the galileon model with global internal symmetries. We argue that the most important feature of the galileon model is that the spacetime indices of galileon's derivatives are appropriately anti-symmetrized and we shall keep this feature when gauging the galileon model. In Section~\ref{ggalf}, we construct the minimal gauge extension of the symmetric galileon model, focusing on the fundamental representations of all the classical groups. We show that its equations of motion remain second order, naturally avoiding Ostrogradski ghosts. This is in contrast to the gravitational covariantization of the galileon model, where Ostrogradski ghosts do arise unless extra terms with non-minimal couplings between the galileon and the curvature tensors are added to the theory. In Section~\ref{ggalc}, we further promote the metric to be dynamical, in which case we show that non-minimal couplings between the galileon and the curvature tensors are required in order to maintain the second order nature of the field equations. In Section~\ref{ggalext}, we discuss non-minimal gauge covariantisations from the global symmetric galileon and gauged galileon models in other representations of the gauge group. We conclude in Section~\ref{summout}.

\section{Galileons with global symmetries} \label{galgb}

First, we briefly review the galileon models with global internal symmetries. The galileon can be identified as the brane bending mode in certain braneworld scenarios. In a braneworld setup with several co-dimensions, the brane has several directions to bend, and the resulting 4D effective theory may contain several galileons. Neglecting the tadpole terms, the general $N$-galileon model in 4D\;\footnote{Through out this paper, we will explicitly work in 4D, but it is clear how to generalize to arbitrary dimensions.} can be written as~\cite{Padilla:2010ir}
\be \label{mltgal}
\mathcal{L} = \sum_{n=2}^{5} \lambda_{i_1 i_2...i_n} \pd^{\mu_2} \pi_{i_1} \pd_{[\mu_2} \pi_{i_2} \pd^{\mu_3} \pd_{\mu_3} \pi_{i_3} ... \pd^{\mu_n} \pd_{\mu_n]} \pi_{i_n}   ,
\ee
where ${}_{i_1}$,${}_{i_2}$,...,${}_{i_n}$ label the different galileons,  $\lambda_{i_1 i_2...i_n}$ are free parameters of the theory and summations over the labeling indices $i_1,...i_n$ are assumed. Note that we define the $n=2$ case as $\lambda_{i_1 i_2} \pd^{\mu_2} \pi_{i_1} \pd_{\mu_2} \pi_{i_2}$. All the down spacetime indices ${}_{\mu_2}$,...,${}_{\mu_n}$ are anti-symmetrized. Since all the up indices are contracted with the down indices, the up indices ${}^{\mu_2}$,...,${}^{\mu_n}$ can also be considered as being anti-symmetrized. Because of this anti-symmetrisation of the spacetime indices, we can freely ``move'' a partial derivative from one $\pi_{i_k}$ to another by integration by parts. This is a key feature that gives the galileon models their interesting properties such as Galilean field symmetry, absence of Ostragradski ghosts and the non-renormalisation theorem (e.g.~\cite{Hinterbichler:2010xn, Deffayet:2010zh}). Another result of the anti-symmetrisation of the spacetime indices is that the indices of $\lambda_{i_1 i_2...i_n}$ can be chosen as symmetric, i.e., $\lambda_{i_1 i_2...i_n}=\lambda_{(i_1 i_2...i_n)}$. By simple combinatorial counting, we can see that the number of free parameters of a general multi-galileon model increases very rapidly with $N$: $\sum^5_{n=2}(N+n-1)!/n!(N-1)!$~\cite{Padilla:2010ir}.

However, if the bulk of the underlying braneworld scenario contains some global symmetry, the resulting galileon theory may inherit this symmetry and the allowed Lagrangian terms become more restricted~\cite{Padilla:2010ir, Hinterbichler:2010xn}. That is, the multiple galileons now form some representation of a Lie group
\be \label{pivector}
\pi = (\pi_1,\pi_2,...,\pi_N)^T  ,
\ee
and, apart form the Galilean field symmetry, the Lagrangian is invariant under a global symmetry transformation
\be
\pi \to e^{i c_a T^a} \pi   ,
\ee
where $c_a$'s are constants across the spacetime, $T^a$'s are the generators of the corresponding representation, and summation over the group indices is assumed. The fundamental representation of a group is arguably the most basic irreducible representation, which defines the corresponding matrix group. Now, if the galileon lives in the fundamental representation\;\footnote{Of course, to construct a Lagrangian invariant for a complex representation, its conjugate also has to be included.} of a classical group SU(N), SO(N) or Sp(N), the galileon Lagrangian is simply given by~\cite{Padilla:2010ir, Hinterbichler:2010xn}
\be \label{frgbsym}
\mathcal{L} = - \pd_\mu \pi^\dagger \pd^\mu \pi
+ \lambda (\pd^\mu \pi^\dagger \pd_{[\mu} \pi ) (\pd^\nu \pd_{\nu} \pi^\dagger \pd^{\rho} \pd_{\rho]} \pi)  ,
\ee
where ${}^\dagger$ is the Hermitian conjugate and $\lambda$ is a free parameter. That is, the theory has only one coupling constant. Mathematically, the Lagrangian is so simple because, in the fundamental representation of the classical groups, the only symmetric primitive invariant of the corresponding Lie algebra which the symmetric galileon lives in is $\delta^i{}_j$~\cite{Fuchs:1997jv}, where the down and up group indices differentiate the fundamental representation and its conjugate respectively. Strictly speaking, for SO(N), there are also $\delta^{ij}$ and $\delta_{ij}$, but the fundamental representation of SO(N) is a real representation, so they can be treated as the same group tensors. For the fundamental representation of the exceptional groups, there may be extra Lagrangian terms. For example, there is another symmetric primitive invariant $d^{ijk}$ in the $E_6$ Lie algebra~\cite{Fuchs:1997jv}, implying that under this symmetry the Lagrangian will contain additionally a cubic term contracted with this $d^{ijk}$.

More generally, the galileon may live in other representations of the internal group. The Lagrangian is more complicated for these cases. For example, in~\cite{Padilla:2010ir}, all possible Lagrangian terms for the adjoint representations of SU(N) and SO(N) were presented. To have a practical taste, all possible Lagrangian terms for the adjoint representation of SO(N) with $N\geq 5$ are given by
\begin{align}
\label{eq:sunAdjoint}
&~
\mathrm{Tr}(\Phi\Box\Phi)
, ~~~~
\mathrm{Tr}(\pd\Phi\pd_[ \Phi\pd\pd_] \Phi)  ,
\nn
&~ \mathrm{Tr}(\pd\Phi\pd_[ \Phi\pd\pd \Phi\pd\pd_] \Phi)
, ~~~~
\mathrm{Tr}(\pd\Phi\pd_[ \Phi)\mathrm{Tr}(\pd\pd \Phi\pd\pd_] \Phi)  ,
\nn
&~
\mathrm{Tr}(\pd\Phi\pd_[ \Phi\pd\pd \Phi\pd\pd \Phi\pd\pd_] \Phi)   ,
\nn
&~
\mathrm{Tr}(\pd\Phi\pd_[ \Phi)\mathrm{Tr}(\pd\pd\Phi \pd\pd \Phi\pd\pd_] \Phi)  ,
\end{align}
where we have suppressed the spacetime indices for simplicity and $\Phi=(\Phi_{i}^{\phantom{i}j})$ is an $N\times N$ traceless hermitian matrix, which lives in the adjoint representation of SU(N), transforming under an SU(N) defining matrix $U=(U_{i}^{\phantom{i}j})_{N\times N}$ as $\Phi \to U \Phi U^{\dagger}$.

\section{Galileons with gauge symmetries} \label{ggalf}

In this section, we promote the internal symmetry of the symmetric galileon to be a local one. That is, we couple the galileon to a gauge field $A_\mu = A_\mu^a T^a$, where $T^a$'s again are the generators of the symmetric galileon's representation.  We will work in the general case where the gauge field is of the Yang-Mills type and the expressions will be written in terms of matrices that live in the representation of the corresponding Lie algebra. The Abelian theory can be obtained as a special case where $A_\mu$ is merely a real function of spacetime\;\footnote{An explicit example of a U(1) gauged galileon model can arise in 3D New Massive Gravity~\cite{arXiv:1103.1351}.}.

We first establish our convention for the gauge field $A_\mu = A_\mu^a T^a$ and the gauge covariant derivative $D_\mu$. We define the field strength as
\begin{align}
\label{gaugefs}
F_{\mu\nu} &= F^c_{\mu\nu} T^c
\\
         &= 2\pd_{[\mu} A_{\nu]} +  i [A_{\mu},A_{\nu}]
\\
         &= 2\pd_{[\mu} A^c_{\nu]}T^c - f^{abc} A^a_\mu A^b_\nu T^c   ,
\end{align}
where $f^{abc}$ are the structure constants of the gauge group satisfying
\be
[T^a,T^b] = i f^{abc} T^c  .
\ee
We are looking for a Lagrangian that is invariant under a gauge transformation
\begin{align}
\pi &\to \Omega(x) \pi = e^{i\alpha_a(x) T^a} \pi    ,
\\
A_\mu &\to \Omega A_\mu \Omega^\dagger + i (\pd_\mu \Omega)\Omega^\dagger    ,
\end{align}
where again $\pi$ is a column vector and $\alpha_a(x)$'s are functions of spacetime. A simple prescription for the gauge covariantisation is to promote the partial derivatives in the galileon theory to gauge covariant derivatives: $\pd_\mu \to D_\mu$. Later, we will also encounter gauge covariant derivatives in curved space. From a viewpoint of differential geometry, we could simply call all these derivatives as the covariant derivative that takes into account all the fibres on a base manifold and changes its form according to the object it acts on. But for concreteness we choose to use different symbols for these different covariant derivatives.

In Minkowski space, the action of the gauge covariant derivative on the galileon $\pi$ is given by
\be
D_\mu \pi= (\pd_\mu + i A_\mu) \pi =(\pd_\mu + i A_\mu^a T^a) \pi   .
\ee
Note that the gauge coupling does not appear in the gauge covariant derivative but will appear in front of the gauge kinetic term. Since the field strength lives in the adjoint representation
\be
F_{\mu\nu}  \to \Omega  F_{\mu\nu} \Omega^\dagger   ,
\ee
the gauge covariant derivative on $F_{\mu\nu}$ is given by
\be
D_\rho F_{\mu\nu} = \pd_\rho F_{\mu\nu} + i [A_{\rho},F_{\mu\nu}]   .
\ee
We will also use the left acting gauge covariant derivatives $\loarrow{D}_{\mu}= (D_\mu)^\dagger$ , whose actons on $\pi^\dagger$ and $F_{\mu\nu}$ are defined by
\begin{align}
\pi^\dagger \loarrow{D}_{\mu} & =\pi^\dagger (\loarrow{\pd}_{\mu} - i A_\mu)= \pd_{\mu} \pi^\dagger - i \pi^\dagger A_\mu
\\
F_{\mu\nu} \loarrow{D}_{\rho} & = {D}_{\rho} F_{\mu\nu}     .
\end{align}

Now, if the galileon lives in the fundamental representation of a classical Lie group SU(N), SO(N) or Sp(N), a direct promotion from $\pd_\mu \to D_\mu$ for the Lagrangian (\ref{frgbsym}) leads to the following Lagrangian
\begin{align} \label{frgal}
\mathcal{L}_{\rm m} &= -\frac{1}{2 g_G^2} \mathrm{tr}(F_{\mu\nu} F^{\mu\nu})   - (D_\mu \pi)^\dagger D^\mu \pi
\nn
&~~~  + \lambda_4 \left((D^\mu \pi)^\dagger D_{[\mu} \pi \right) \left( (D^\nu D_{\nu} \pi)^\dagger D^{\rho} D_{\rho]} \pi \right)
\nn
& =  -\frac{1}{2 g_G^2} \mathrm{tr}(F_{\mu\nu} F^{\mu\nu})  - \pi^\dagger \loarrow{D}_\mu D^\mu \pi
\nn
&~~~ + \lambda_4 \left(\pi^\dagger \loarrow{D}^\mu D_{[\mu} \pi \right) \left( \pi^\dagger \loarrow{D}_{\nu} \loarrow{D}^{\nu}  D^{\rho} D_{\rho]} \pi \right)   ,
\end{align}
where $g_G$ is the gauge coupling constant and $\lambda_4$ may be called the galileon coupling constant. Note that in the Lagrangian (\ref{frgal}) both the up and down indices $\mu,~\nu,~\rho$ in the quartic $\pi$ term are anti-symmetrized. We will see that the equations of motion of this Lagrangian does not contain higher order derivatives. However, the Lagrangian (\ref{frgal}), which we will refer to as the minimal gauging case, is not the only possible way to gauge the symmetric galileon model (\ref{frgbsym}). We will discuss this later in Section~\ref{ggalext}. Our conclusion remains the same if all the possible ways of gauging the Lagrangian (\ref{frgbsym}) are considered.

To facilitate the derivation of $\pi_i$'s and $A^\sigma$'s equations of motion, we establish a few further relations regarding the gauge covariant derivative. The commutators of $D_\mu$ and $\loarrow{D}_\mu$ with themselves are
\be \label{gdcom}
 [D_\mu, D_\nu] \pi = i F_{\mu\nu} \pi,
\quad \pi^\dagger [\loarrow{D}_\mu, \loarrow{D}_\nu]
   = i \pi^\dagger F_{\mu\nu}   .
\ee
Note that when acting on the field strength, the commutator of gauge covariant derivatives obeys a different rule:
$[D_\mu, D_\nu] F_{\rho\sigma} = i [F_{\mu\nu},F_{\rho\sigma}]$. To derive an equation of motion, we need to integrate by parts and use the Leibniz rule. In Minkowski space, this is with respect to the partial derivative and assumes integration over the total derivative $\pd_\mu (...)^\mu$ vanish. In a gauge theory, the natural derivatives appearing in the action are the gauge covariant derivative. Nevertheless, as one may expect, it can be shown that the usual rules of derivatives apply for the gauge covariant derivative. Particularly, we have the following relations
\be
\int\! \ud^4 x  (\eta^\dagger \loarrow{D}_\mu) \psi^\mu (\phi^\dagger \varphi)
=
-\int\! \ud^4 x  \eta^\dagger D_\mu (\psi^\mu (\phi^\dagger \varphi))    ,
\ee
and
\be
 {D}_\mu ( \psi^\mu (\phi^\dagger \varphi) )
 =
 {D}_\mu \psi^\mu  (\phi^\dagger \varphi)
 +  \psi^\mu (\phi^\dagger \loarrow{D}_\mu \varphi)
 + \psi^\mu (\phi^\dagger D_\mu \varphi)    ,
\ee
where $\eta$, $\psi$, $\phi$ and $\varphi$ take values on the same representation as $\pi$ (a column vector in matrix form) but may carry gauge covariant derivatives and properly contracted spacetime indices. There are also similar relations where the up ${}^\mu$ index is carried by $\eta$, $\phi$ or $\varphi$. As an aside, when $\eta$ and $\psi$ live in a representation different from that of $\phi$ and $\varphi$, these relations still hold, provided the gauge covariant derivatives acting on these group vectors are defined appropriately in the corresponding representations. In short, we can perform integration by parts and use the Leibniz rule for gauge covariant derivatives in a way similar to that of the partial derivatives.

With these rules established, we can easily derive the equation of motion for $\pi^\dagger$ ($\pi$'s equation of motion is the hermitian conjugate of $\pi^\dagger$'s). For a simpler presentation of the results, we define
\be \label{Pidef}
\Pi^{\mu}{}_{\nu}{}^{...}_{...}= D^\mu D_\nu \cdots \pi ,
\quad~~~
\loarrow{\Pi}^{...}_{...}{}^{\mu}{}_{\nu} = \pi^\dagger \cdots \loarrow{D}^\mu \loarrow{D}_\nu  .
\ee
In matrix form, $\pi^\dagger$'s equation of motion is given by
\be
\mathcal{E}^{\pi^\dagger} = \mathcal{E}_2^{\pi^\dagger} + \lambda_4 \mathcal{E}_4^{\pi^\dagger}  ,
\ee
where
\be
\mathcal{E}_2^{\pi^\dagger} = \Pi{}^\mu{}_\mu   ,
\ee
and
\begin{align}
 \mathcal{E}_4^{\pi^\dagger} & =
- 2\Pi^{\mu}{}_{[\mu} \loarrow{\Pi}_{\nu}{}^{\nu} \Pi{}^{\rho}{}_{\rho]}
- \Pi_{[\mu} \loarrow{\Pi}_{\nu}{}^{\nu}{}^{\mu} \Pi{}^{\rho}{}_{\rho]}
- \Pi_{[\mu} \loarrow{\Pi}_{\nu}{}^{\nu} \Pi{}^{\mu}{}^{\rho}{}_{\rho]}
\nn
&~~+\Pi{}_{[\nu}{}^\nu{}^\rho{}_\rho \loarrow{\Pi}{}^{\mu} \Pi{}_{\mu]}
+\Pi{}^\nu{}^{\rho}{}_{[\rho} \loarrow{\Pi}{}^{\mu}{}_{\nu} \Pi{}_{\mu]}
+\Pi{}^\nu{}^{\rho}{}_{[\rho} \loarrow{\Pi}{}^{\mu} \Pi{}_{\nu}{}_{\mu]}
\nn
&~~+\Pi{}_{[\nu}{}^{\rho}{}_{\rho} \loarrow{\Pi}{}^{\mu}{}^{\nu} \Pi{}_{\mu]}
+\Pi{}^{\rho}{}_{[\rho} \loarrow{\Pi}{}^{\mu}{}^{\nu}{}_{\nu} \Pi{}_{\mu]}
+\Pi{}^{\rho}{}_{[\rho} \loarrow{\Pi}{}^{\mu}{}^{\nu} \Pi{}_{\nu}{}_{\mu]}
\nn
&~~+\Pi{}_{[\nu}{}^{\rho}{}_{\rho} \loarrow{\Pi}{}^{\mu} \Pi{}^{\nu}{}_{\mu]}
+\Pi{}^{\rho}{}_{[\rho} \loarrow{\Pi}{}^{\mu} \Pi{}_{\nu}{}^{\nu}{}_{\mu]}   .
\end{align}
There are terms with 3 or 4 gauge covariant derivatives acting on the galileon, but, thanks to the anti-symmetrisation of the spacetime indices, all the higher order derivatives exactly cancel each other. To see this clearly, we use  Eqs.~(\ref{gdcom}), which leads to
\begin{align}
 \mathcal{E}_4^{\pi^\dagger} & =
- 2\Pi^{\mu}{}_{[\mu} \loarrow{\Pi}_{\nu}{}^{\nu} \Pi{}^{\rho}{}_{\rho]}
- \frac{i}{2} \Pi_{[\mu} \loarrow{\Pi}{}_{\nu}F{}^{\nu\mu} \Pi{}^{\rho}{}_{\rho]}
\nn
&~~- \frac{i}{2} \Pi_{[\mu} \loarrow{\Pi}_{\nu}{}^{\nu} F{}^{\mu\rho} \Pi{}_{\rho]}
+ \frac{i}{2} (D_{[\nu}F{}^{\nu\rho})\Pi{}_\rho \loarrow{\Pi}{}^{\mu} \Pi{}_{\mu]}
\nn
&~~ - \frac14 F{}^{\nu\rho}F_{[\nu\rho}\pi \loarrow{\Pi}{}^{\mu} \Pi{}_{\mu]}
+\frac{i}{2} F{}^{\nu\rho}\Pi{}_{[\rho} \loarrow{\Pi}{}^{\mu}{}_{\nu} \Pi{}_{\mu]}
\nn
&~~-\frac14 F{}^{\nu\rho}\Pi{}_{[\rho} \loarrow{\Pi}{}^{\mu} F_{\nu\mu]}\pi
-\frac14 F{}_{[\nu\rho}\Pi{}^{\rho} \pi^\dagger F^{\mu\nu} \Pi{}_{\mu]}
\nn
&~~-\frac{i}{2}D_{[\nu} (F^\mu{}_\rho \pi) \pi^\dagger F^{\rho\nu} \Pi{}_{\mu]}
+\frac{i}{2}\Pi{}^{\rho}{}_{[\rho} (\pi^\dagger F^{\mu\nu})\loarrow{D}_{\nu} \Pi{}_{\mu]}
\nn
&~~-\frac14 \Pi{}^{\rho}{}_{[\rho} \pi^\dagger F^{\mu\nu} F{}_{\nu\mu]}\pi
+\frac{i}{2}F_{[\nu\rho}\Pi{}^{\rho} \loarrow{\Pi}{}^{\mu} \Pi{}^{\nu}{}_{\mu]}
\nn
&~~-\frac{i}{2}D_{[\nu} (F^\mu{}_\rho\pi) \loarrow{\Pi}^{\rho}\Pi{}^\nu{}_{\mu]}
+\frac{i}{2}\Pi{}^{\rho}{}_{[\rho} \loarrow{\Pi}{}^{\mu} D^{\nu}(F_{\nu\mu]}\pi)
\nn
&~~-\frac{i}{2}\Pi{}^{\rho}{}_{[\rho} \loarrow{\Pi}{}^{\nu} F{}_{\nu}{}^{\mu}\Pi{}_{\mu]}   .
\end{align}
Since $F_{\mu\nu}$ only contains first order derivatives, $\pi^\dagger$'s equation of motion does not have terms with higher than second order derivatives.

Next, we also need to show that $A^\sigma$'s equation of motion does not have higher order derivatives. In deriving $A^\sigma$'s equation of motion, the following corresponding relations are useful,
\begin{align}
{\sf Lagran.~variation}
&\phantom{\xrightarrow{\mathrm{EoM}~\mathrm{of}~A^\sigma}}
{\sf ~~EoM~terms~produced}
\nn
\label{geom1}
\phi^\dagger_\mu \delta_A({D}^\mu \pi)
&\xrightarrow{\mathrm{EoM}~\mathrm{of}~A^\sigma}
i \pi \phi^\dagger_\sigma  ,
\\
\label{geom2}
\delta_A(\pi^\dagger \loarrow{D}^\mu) \phi_\mu
&\xrightarrow{\mathrm{EoM}~\mathrm{of}~A^\sigma}
-i \phi_\sigma \pi^\dagger  ,
\\
\label{geom3}
\psi^\dagger{}^\mu{}_\nu \delta_A({D}^\nu D_\mu \pi)
&\xrightarrow{\mathrm{EoM}~\mathrm{of}~A^\sigma}
i D_\mu \pi \psi^\dagger{}^\mu{}_\sigma
-i\pi \psi^\dagger{}_\sigma{}_\nu \loarrow{D}^\nu  ,
\\
\label{geom4}
\delta_A(\pi^\dagger \loarrow{D}_\mu \loarrow{D}^\nu) \psi{}^\mu{}_\nu
&\xrightarrow{\mathrm{EoM}~\mathrm{of}~A^\sigma}
\!-i \psi^\mu{}_\sigma \pi^\dagger \loarrow{D}_\mu
+i  D^\nu \psi_\sigma{}_\nu \pi^\dagger \! ,
\end{align}
where $\delta_A$ means variation with respect to $A^\sigma$, $\phi_\mu$ and $\psi^\mu{}_\nu$ live in the same representation as $\pi$, and the RHS of the arrow is the equation of motion terms arisen from the Lagrangian variation of the LHS. These correspondences can be checked by explicit calculations. Using these relations, as well as the commutator relation (\ref{gdcom}), the equation of motion for $A^\sigma$ in matrix form is given by
\be
\mathcal{E}^{A^\sigma} = \mathcal{E}^{A^\sigma}_F + \mathcal{E}^{A^\sigma}_2 + \lambda_4 \mathcal{E}^{A^\sigma}_4
\ee
where
\be
\mathcal{E}^{A^\sigma}_F = \frac{2}{g_G^2} D^\rho F_{\rho\sigma}
\quad
\mathcal{E}^{A^\sigma}_2 = -i \Pi_\sigma \pi^\dagger + i\pi \loarrow{\Pi}_\sigma
\ee
and
\begin{align}
\mathcal{E}^{A^\sigma}_4 &=
-i \Pi{}_{[\sigma} \pi^\dagger \loarrow{\Pi}{}_\nu{}^\nu \Pi{}^\rho{}_{\rho]}
+ i \eta_{\sigma[\mu}\pi\loarrow{\Pi}{}^\mu \loarrow{\Pi}{}_\nu{}^\nu \Pi{}^\rho{}_{\rho]}
\nn&~~
-i\Pi{}^{\rho}{}_{[\rho} \loarrow{\Pi}{}_\sigma \loarrow{\Pi}{}^\mu \Pi_{\mu]}
-\frac{1}{2}\eta_{\sigma[\nu}F{}^{\nu\rho}\Pi{}_{\rho} \pi^\dagger \loarrow{\Pi}{}^\mu \Pi_{\mu]}
\nn&~~
-\frac{1}{2}\eta_{\sigma[\nu}\Pi{}^{\rho}{}_{\rho} \pi^\dagger \pi^\dagger F{}^{\mu\nu} \Pi_{\mu]}
+i\eta_{\sigma[\nu} \Pi{}^{\rho}{}_{\rho} \pi^\dagger \loarrow{\Pi}{}^\mu \Pi^\nu{}_{\mu]}
\nn&~~
+i\Pi_{[\sigma} \loarrow{\Pi}^\nu{}_\nu \loarrow{\Pi}^\mu \Pi{}_{\mu]}
+\frac{1}{2}\eta_{\sigma[\rho} \loarrow{\Pi}_\nu F^{\nu\rho} \loarrow{\Pi}^\mu \Pi_{\mu]}
\nn&~~
-\frac{1}{2}\eta_{\sigma[\rho} \pi (\pi^\dagger  F^{\rho}{}_\nu) \loarrow{D}^\nu \loarrow{\Pi}^\mu \Pi_{\mu]}
+\frac{1}{2}\eta_{\sigma[\rho} \pi \loarrow{\Pi}^\nu{}_\nu \pi^\dagger F^{\mu\rho} \Pi_{\mu]}
\nn&~~
-i\eta_{\sigma[\rho} \pi \loarrow{\Pi}^\nu{}_\nu \loarrow{\Pi}^\mu \Pi^\rho{}_{\mu]}
\end{align}

Therefore, none of the equations of motion for the theory (\ref{frgal}) has higher than second order derivatives. In other words, the galileon model can be naturally extended to couple to gauge fields without introducing Ostrogradski ghosts. As we will discuss in Section~\ref{ggalext}, the same conclusion holds for any representation of the gauge group in Minkowski space.

\section{Gauged galileons in a curved background} \label{ggalc}

We now allow the spacetime metric to fluctuate away from Minkowski space and construct a gauged galileon field theory in this new curved space background. This may be related to a braneworld scenario where the brane has non-trivial intrinsic geometry.

At an arbitrary point of the curved spacetime, there is a tangent space as well as a Lie algebra of the gauge group. But they are different fibres on the base manifold, so the covariantisation of the galileon theory with gauge symmetries obeys the ``comma-goes-to-semicolon'' rule~\cite{weinberg}:
\begin{align}
D_\mu \psi_\nu= (\pd_\mu + i A_\mu)\psi_\nu
\to
\mathcal{D}_\mu \psi_\nu = (\nabla_\mu + i A_\mu)  \psi_\nu   ,
\end{align}
where $\nabla_\mu$ is the (tangent space) covariant derivative multiplied by the identity matrix of the corresponding representation of the gauge Lie algebra. Now, the commutators of $\mathcal{D}_\mu$ and $\loarrow{\mathcal{D}}_\mu = (\mathcal{D}_\mu)^\dagger$ are given by
\begin{align}
\label{comcs1}
[\mathcal{D}_\mu, \mathcal{D}_\nu] \psi^\sigma &=
i F_{\mu\nu} \psi^\sigma + R^{\sigma}{}_\rho{}_{\mu\nu} \psi^\rho    ,
\\
\label{comcs2}
\psi^\dagger{}^\sigma [\loarrow{\mathcal{D}}_\mu, \loarrow{\mathcal{D}}_\nu]  &=
i\psi^\dagger{}^\sigma F_{\mu\nu} -  \psi^\dagger{}^\rho R^{\sigma}{}_\rho{}_{\mu\nu}   ,
\end{align}
where $\psi^\sigma$ lives in the same gauge representation as $\pi$ and $R^{\sigma}{}_\rho{}_{\mu\nu}$ is the Riemann tensor of the curved spacetime. We have assumed the underlying gravity theory is torsion free.  Note that the Riemann tensor term does not arise when acting the commutator on the spacetime scalars $\pi$ or $\pi^\dagger$.

In the fundamental representation of the classical groups, a direct use of the ``comma-goes-to-semicolon'' rule to the Lagrangian (\ref{frgal}) leads to the Lagrangian
\begin{align} \label{frgalcs1}
\mathcal{L}^{\rm CS}_{\rm m} &= \sqrt{-g} \left[ R -\frac{1}{2 g_G^2} \mathrm{tr}(F_{\mu\nu} F^{\mu\nu})   - (\mathcal{D}_\mu \pi)^\dagger \mathcal{D}^\mu \pi  \right.
\nn
&~~~ \left. + \gamma_4 \left((\mathcal{D}^\mu \pi)^\dagger \mathcal{D}_{[\mu} \pi \right) \left( (\mathcal{D}^\nu \mathcal{D}_{\nu} \pi)^\dagger \mathcal{D}^{\rho} \mathcal{D}_{\rho]} \pi \right) \phantom{\frac11}\!\!\!\!\! \right] ,
\end{align}
where $R$ is the Ricci scalar and $\gamma_4$ is the galileon coupling in curved space. But this Lagrangian actually leads to a higher than second order equation of motion for $\pi^\dagger$, similar to the case of the covariant galileon studied in~\cite{Deffayet:2009wt}. Specifically, there is one higher order derivative term appearing in $\pi^\dagger$'s equation of motion, namely,
\be
\frac{1}{2}(\nabla_{[\nu}R_{\rho}{}^\lambda{}^{\nu\rho})\Xi{}_{|\lambda|} \loarrow{\Xi}{}^{\mu} \Xi{}_{\mu]}
=
\frac16 \nabla^{\rho}G_{\mu\nu}\Xi{}_{\rho} \loarrow{\Xi}{}^{\mu} \Xi{}^{\nu}    ,
\ee
where we have used Bianchi's identities, $G_{\mu\nu}=R_{\mu\nu}-Rg_{\mu\nu}/2$, and, analogous to Eq.~(\ref{Pidef}), we have defined
\be
\Xi^{\mu}{}_{\nu}{}^{...}_{...}= \mathcal{D}^\mu \mathcal{D}_\nu \cdots \pi ,
\quad~~~
\loarrow{\Xi}^{...}_{...}{}^{\mu}{}_{\nu} = \pi^\dagger \cdots \loarrow{\mathcal{D}}^\mu \loarrow{\mathcal{D}}_\nu   .
\ee
So the Lagrangian (\ref{frgalcs1}) is not the one we are looking for. We want a Lagrangian whose equations of motion contain at most second order derivatives and reduce to the Lagrangian (\ref{frgal}) in the flat space approximation. This Lagrangian is given by
\be \label{frgalcs}
\bar{\mathcal{L}}^{\rm CS}_{\rm m} = \mathcal{L}^{\rm CS}_{\rm m} + \mathcal{L}^{\rm CS}_{\rm C}  ,
\ee
where $\mathcal{L}^{\rm CS}_{\rm C} $ is a coupling between the Einstein tensor and the galileon
\be \label{cnterm}
\mathcal{L}^{\rm CS}_{\rm C}    =
\frac{\gamma_4}{6}\sqrt{-g} G_{\mu\nu}\loarrow{\Xi}^{\rho}\Xi{}_{\rho} \loarrow{\Xi}{}^{\mu} \Xi{}^{\nu}  .
\ee
Now, the modified Lagrangian (\ref{frgalcs}) has a well-posed Cauchy problem, as we are about to check explicitly.

The equation of motion for $\pi^\dagger$ is, in matrix form,
\be
\mathcal{X}^{\pi^\dagger} = \mathcal{X}_2^{\pi^\dagger} + \gamma_4 \mathcal{X}_4^{\pi^\dagger}   ,
\ee
where
\be
\mathcal{X}_2^{\pi^\dagger} = \Xi{}^\mu{}_\mu  ,
\ee
and
\begin{align}
\mathcal{X}_4^{\pi^\dagger} & =
- 2\Xi^{\mu}{}_{[\mu} \loarrow{\Xi}_{\nu}{}^{\nu} \Xi{}^{\rho}{}_{\rho]}
- \frac{i}{2}\Xi_{[\mu} \loarrow{\Xi}{}_{\nu}F{}^{\nu\mu} \Xi{}^{\rho}{}_{\rho]}
\nn
&~~-\frac12 \Xi_{[\mu} \loarrow{\Xi}{}_{|\lambda|}R_\nu{}^{\lambda}{}^{\nu\mu} \Xi{}^{\rho}{}_{\rho]}
- \frac{i}{2}\Xi_{[\mu} \loarrow{\Xi}_{\nu}{}^{\nu} F{}^{\mu\rho} \Xi{}_{\rho]}
\nn
&~~+\frac12 \Xi_{[\mu} \loarrow{\Xi}_{\nu}{}^{\nu} R_{\rho]}{}^{\lambda\mu\rho} \Xi{}_{\lambda}
+ \frac{i}{2} (\mathcal{D}_{[\nu}F{}^{\nu\rho})\Xi{}_\rho \loarrow{\Xi}{}^{\mu} \Xi{}_{\mu]}
\nn
&~~ -\frac16 G_{\mu\nu} \mathcal{D}^{\rho}(\Xi{}_{\rho} \loarrow{\Xi}{}^{\mu} \Xi{}^{\nu})
-\frac16 G_{\mu\nu} \mathcal{D}^{\mu}(\Xi{}^{\nu} \loarrow{\Xi}{}^{\rho} \Xi{}_{\rho})
\nn
&~~ +\frac{i}{2} F{}^{\nu\rho}\Xi{}_{[\rho} \loarrow{\Xi}{}^{\mu}{}_{\nu} \Xi{}_{\mu]}
+ \frac12 R_{[\rho}{}^{\lambda\nu\rho}\Xi_{|\lambda|} \loarrow{\Xi}{}^{\mu}{}_{\nu} \Xi{}_{\mu]}
\nn
&~~-\frac14 F{}^{\nu\rho}\Xi{}_{[\rho} \loarrow{\Xi}{}^{\mu} F_{\nu\mu]}\pi
+\frac{i}4  R_{[\rho}{}^{\lambda\nu\rho}\Xi_{|\lambda|} \loarrow{\Xi}{}^{\mu} F_{\nu\mu]}\pi
\nn
&~~-\frac14  F{}_{[\nu\rho}\Xi{}^{\rho} \pi^\dagger F^{\mu\nu} \Xi{}_{\mu]}
+\frac{i}4  R^\rho{}_{\lambda[\nu\rho}\Xi{}^{\lambda} \pi^\dagger F^{\mu\nu} \Xi{}_{\mu]}
\nn
&~~-\frac{i}4 \mathcal{D}_{[\nu} (F^\mu{}_\rho \pi) \pi^\dagger F^{\rho\nu} \Xi{}_{\mu]}
+\frac{i}2  \Xi{}^{\rho}{}_{[\rho} (\pi^\dagger F^{\mu\nu})\loarrow{\mathcal{D}}_{\nu} \Xi{}_{\mu]}
\nn
&~~-\frac14  \Xi{}^{\rho}{}_{[\rho} \pi^\dagger F^{\mu\nu} F{}_{\nu\mu]}\pi
+\frac{i}2   F_{[\nu\rho}\Xi{}^{\rho} \loarrow{\Xi}{}^{\mu} \Xi{}^{\nu}{}_{\mu]}
\nn
&~~+\frac12   R^\rho{}_{\lambda[\nu\rho}\Xi{}^{\lambda} \loarrow{\Xi}{}^{\mu} \Xi{}^{\nu}{}_{\mu]}
-\frac{i}2     \mathcal{D}_{[\nu} (F^\mu{}_\rho \pi) \loarrow{\Xi}^{\rho}\Xi{}^\nu{}_{\mu]}
\nn
&~~+\frac{i}2    \Xi{}^{\rho}{}_{[\rho} \loarrow{\Xi}{}^{\mu} \mathcal{D}^{\nu}(F_{\nu\mu]}\pi)
-\frac{i}2      \Xi{}^{\rho}{}_{[\rho} \loarrow{\Xi}{}^{\nu} F{}_{\nu}{}^{\mu}\Xi{}_{\mu]}
\nn
&~~-\frac12   \Xi{}^{\rho}{}_{[\rho} \loarrow{\Xi}{}^{\nu} R{}_{\mu}{}^\lambda{}_{\nu]}{}^{\mu} \Xi{}_{\lambda}
- \frac14    F{}^{\nu\rho}F_{[\nu\rho}\pi \loarrow{\Xi}{}^{\mu} \Xi{}_{\mu]}  .
\end{align}
Note that the index ${}_\lambda$ is not in the anti-symmetrisation. We can see that there are no higher order derivatives in $\pi^\dagger$'s equation of motion.

We also have to check the equations of motion for $A^\sigma$ and $g^{\alpha\beta}$. The remarkable thing about the (counter) term (\ref{cnterm}) is that it also reduces all the higher derivatives in the equations of motion for $A^\sigma$ and $g^{\alpha\beta}$.

By explicit calculations, we can also establish a set of rules to facilitate the derivation of the equation of motion for $A^\sigma$, which are equivalent to those of~(\ref{geom1}-\ref{geom4}), with $D_\mu$ replaced by $\mathcal{D}_\mu$. Using these relations, for the modified Lagrangian (\ref{frgalcs}), $A^\sigma$'s equation of motion is given by
\be
\mathcal{X}^{A^\sigma} = \mathcal{X}^{A^\sigma}_F + \mathcal{X}^{A^\sigma}_2 + \gamma_4 \mathcal{X}^{A^\sigma}_4
\ee
where
\be
\mathcal{X}^{A^\sigma}_F = \frac{2}{g_G^2} \mathcal{D}^\rho F_{\rho\sigma}
\quad
\mathcal{X}^{A^\sigma}_2 = -i \Xi_\sigma \pi^\dagger + i\pi \loarrow{\Xi}_\sigma  ,
\ee
and
\begin{align}
\mathcal{X}^{A^\sigma}_4 &=
-i \Xi{}_{[\sigma} \pi^\dagger \loarrow{\Xi}{}_\nu{}^\nu \Xi{}^\rho{}_{\rho]}
+ i g_{\sigma[\mu}\pi\loarrow{\Xi}{}^\mu \loarrow{\Xi}{}_\nu{}^\nu \Xi{}^\rho{}_{\rho]}
\nn&~~
-i\Xi{}^{\rho}{}_{[\rho} \loarrow{\Xi}{}_\sigma \loarrow{\Xi}{}^\mu \Xi_{\mu]}
-\frac12   g_{\sigma[\nu}F{}^{\nu\rho}\Xi{}_{\rho} \pi^\dagger \loarrow{\Xi}{}^\mu \Xi_{\mu]}
\nn&~~
+\frac{i}2   g_{\sigma[\nu}R_{\rho}{}^{\lambda\nu\rho}\Xi{}_{|\lambda|} \pi^\dagger \loarrow{\Xi}{}^\mu \Xi_{\mu]}
-\frac12    g_{\sigma[\nu}\Xi{}^{\rho}{}_{\rho} \pi^\dagger \pi^\dagger F{}^{\mu\nu} \Xi_{\mu]}
\nn&~~
+ig_{\sigma[\nu} \Xi{}^{\rho}{}_{\rho} \pi^\dagger \loarrow{\Xi}{}^\mu \Xi^\nu{}_{\mu]}
+i\Xi_{[\sigma} \loarrow{\Xi}^\nu{}_\nu \loarrow{\Xi}^\mu \Xi{}_{\mu]}
\nn&~~
+\frac12  g_{\sigma[\rho} \loarrow{\Xi}_\nu F^{\nu\rho} \loarrow{\Xi}^\mu \Xi_{\mu]}
-\frac{i}2  g_{\sigma[\rho} \loarrow{\Xi}_{|\lambda|} R_\nu{}^{\lambda}{}^{\nu\rho} \loarrow{\Xi}^\mu \Xi_{\mu]}
\nn&~~
-\frac12  g_{\sigma[\rho} \pi (\pi^\dagger  F^{\rho}{}_\nu) \loarrow{\mathcal{D}}^\nu \loarrow{\Xi}^\mu \Xi_{\mu]}
+\frac12  g_{\sigma[\rho} \pi \loarrow{\Xi}^\nu{}_\nu \pi^\dagger F^{\mu\rho} \Xi_{\mu]}
\nn&~~
-ig_{\sigma[\rho} \pi \loarrow{\Xi}^\nu{}_\nu \loarrow{\Xi}^\mu \Xi^\rho{}_{\mu]}
-\frac{i}6  G_{\mu\nu} \Xi{}_{\sigma} \pi^\dagger \loarrow{\Xi}{}^{\mu} \Xi{}^{\nu}
\nn&~~
+\frac{i}6  G_{\mu\nu}\pi \loarrow{\Xi}_{\sigma} \loarrow{\Xi}{}^{\mu} \Xi{}^{\nu}
-\frac{i}6  G_{\sigma\nu}  \Xi{}^{\nu} \pi^\dagger \loarrow{\Xi}^{\rho}\Xi{}_{\rho}
\nn&~~
+\frac{i}6  G_{\mu\sigma} \pi \loarrow{\Xi}{}^{\mu} \loarrow{\Xi}^{\rho}\Xi{}_{\rho}  .
\end{align}
Again, the index ${}_\lambda$ is not in the anti-symmetrisation, and there are no higher order derivative terms.

Finally, we show that the equation of motion for the metric $g^{\alpha\beta}$ does not contain higher order derivatives. The full equation of motion for $g^{\alpha\beta}$ is lengthy, and we do display it here. Instead, since we are only concerned whether there are higher order derivatives, it is suffice to only focus on the potentially dangerous terms, i.e., terms potentially having higher order derivatives, and check whether they indeed cancel each other. First, note that the conventional Lagrangian terms
\be
\sqrt{-g}R,
\quad - \frac{\sqrt{-g}}{2 g_G^2} \mathrm{tr}(F_{\mu\nu} F^{\mu\nu}) ,
\quad
 -\sqrt{-g} \loarrow{\Xi}^\mu \Xi_{\mu}
\ee
do not give rise to equation of motion terms with higher order derivatives. To derive $g^{\alpha\beta}$'s equation of motion for the remaining terms, note that the variations with respect to the metric $g^{\alpha\beta}$ and $\sqrt{-g}$ (rather than derivatives of $g^{\alpha\beta}$) do not give rise to higher order derivative terms. The only potential higher order derivatives come from the following variations
\begin{align}
\label{hod1}
\delta\mathcal{L}^{\rm H}_{4}&=
-\gamma_4\sqrt{-g}\loarrow{\Xi}^\mu \Xi_{[\mu} \big(\delta \Gamma^\tau_{\nu|\sigma} g^{\sigma\nu} \loarrow{\Xi}_{\tau|} \Xi^\rho{}_{\rho]}
\nn
&\qquad\qquad\qquad~~~~~~~~~~~~~+\loarrow{\Xi}_\nu{}^\nu \delta \Gamma^\tau_{\rho]\sigma} g^{\sigma\rho} {\Xi}_{\tau} \big)
\\
\label{hod2}
\delta\mathcal{L}^{\rm H}_{\rm C}&=\frac{\gamma_4}{6} \sqrt{-g}(\delta R_{\mu\nu}\!-\!\frac12 g^{\alpha\beta}\delta R_{\alpha\beta}g_{\mu\nu})\loarrow{\Xi}^{\rho}\Xi{}_{\rho} \loarrow{\Xi}{}^{\mu} \Xi{}^{\nu}   .
\end{align}
where $\delta \Gamma^\tau_{\rho\sigma}$ is the variation of the Levi-Civita symbol with respect to $g^{\alpha\beta}$ (including $g^{\alpha\beta}$'s derivatives) and the down indices ${}_\sigma$ and ${}_\tau$ are not in the anti-symmetrisation. We will see that both of these two variations give rise to genuine higher order derivative terms, which exactly cancel each other.

Neglecting the terms that obviously do not contain higher order derivatives, the equation of motion from Eq.~(\ref{hod1}) includes
\begin{align}
\mathcal{X}_4^{g^{\alpha\beta}} & \supset
\frac12 g_{\alpha[\beta}\loarrow{\Xi}^\mu \Xi_{\mu}\loarrow{\Xi}_{|\tau|} \Xi^{\tau\rho}{}_{\rho]}
+\frac12 g_{\alpha[\beta}\loarrow{\Xi}^\mu \Xi_{\mu}\loarrow{\Xi}_{\nu]}{}^{\nu\tau} \Xi_{\tau}
\nn&~~
-\frac12 g_{\alpha[\nu}\loarrow{\Xi}^\mu \Xi_{\mu}\loarrow{\Xi}_{|\beta|} \Xi^{\nu\rho}{}_{\rho]}
-\frac12 \loarrow{\Xi}^\mu \Xi_{[\mu}\loarrow{\Xi}_{|\beta|} \Xi_\alpha{}^{\rho}{}_{\rho]}
\nn&~~
-\frac12 g_{\alpha[\rho}\loarrow{\Xi}^\mu \Xi_{\mu}\loarrow{\Xi}_{\nu]}{}^{\nu\rho} \Xi_\beta
-\frac12 \loarrow{\Xi}^\mu \Xi_{[\mu}\loarrow{\Xi}_{\nu}{}^{\nu}{}_{\alpha]} \Xi_\beta  .
\end{align}
The last 4 terms seem to have third order derivatives, but they are protected by the anti-symmetrisation of the spacetime indices. For each of these terms, two indices of the third order derivative are anti-symmetrized, thus we can use the commutator relations~(\ref{comcs1}-\ref{comcs2}) to reduce them to lower derivative terms involving the gauge field strength or the curvature tensors. Therefore, for the variation (\ref{hod1}), the only non-reducible higher order derivatives are contained in
\be \label{eomcs4}
\mathcal{X}_4^{g^{\alpha\beta}}  \supset
\frac12 g_{\alpha[\beta}\loarrow{\Xi}^\mu \Xi_{\mu}\loarrow{\Xi}_{|\tau|} \Xi^{\tau\rho}{}_{\rho]}
+\frac12 g_{\alpha[\beta}\loarrow{\Xi}^\mu \Xi_{\mu}\loarrow{\Xi}_{\nu]}{}^{\nu\tau} \Xi_{\tau}  .
\ee

On the other hand, neglecting the terms that do not contain apparent higher order derivatives, the equation of motion terms from the variation~(\ref{hod2}) include
\begin{align}
\mathcal{X}_{\rm C}^{g^{\alpha\beta}}   & \supset
-\frac{1}{12} g_{\alpha\beta} \loarrow{\Xi}^{\mu} \Xi_{\mu} \loarrow{\Xi}^{\nu} \Xi_{\rho}{}^{\rho}{}_{\nu}
+\frac{1}{12} g_{\alpha\beta} \loarrow{\Xi}^{\mu} \Xi^{\nu} \loarrow{\Xi}^{\rho} \Xi_{\mu\rho\nu}
\nn&~~
-\frac{1}{12} \loarrow{\Xi}_{\alpha} \Xi^{\nu} \loarrow{\Xi}^{\rho} \Xi_{\beta\rho\nu}
+\frac{1}{12} \loarrow{\Xi}_{\alpha} \Xi_{\beta} \loarrow{\Xi}^{\rho} \Xi_{\mu}{}^\mu{}_\rho
\nn&~~
-\frac{1}{12} \loarrow{\Xi}^{\mu} \Xi_{\alpha} \loarrow{\Xi}^{\rho} \Xi_{\beta\rho\mu}
+\frac{1}{12} \loarrow{\Xi}^{\mu} \Xi_{\mu} \loarrow{\Xi}^{\nu} \Xi_{\alpha\nu\beta}
\nn&~~
-\frac{1}{12}g_{\alpha\beta} \loarrow{\Xi}^{\mu} \Xi_{\mu} \loarrow{\Xi}_{\rho}{}^\rho{}_\nu \Xi^{\nu}
+\frac{1}{12}g_{\alpha\beta} \loarrow{\Xi}^{\mu} \Xi^{\nu} \loarrow{\Xi}_{\mu\rho\nu} \Xi^{\rho}
\nn&~~
-\frac{1}{12} \loarrow{\Xi}_{\alpha} \Xi^{\nu} \loarrow{\Xi}_{\beta\rho\nu} \Xi^{\rho}
+\frac{1}{12} \loarrow{\Xi}_{\alpha} \Xi_{\beta} \loarrow{\Xi}_{\nu}{}^\nu{}_\rho \Xi^{\rho}
\nn&~~
-\frac{1}{12} \loarrow{\Xi}^{\mu} \Xi_{\alpha} \loarrow{\Xi}_{\beta\rho\mu} \Xi^{\rho}
+\frac{1}{12} \loarrow{\Xi}^{\nu} \Xi_{\nu} \loarrow{\Xi}_{\alpha\rho\beta} \Xi^{\rho}
\nn&~~
-\frac{1}{12}g_{\alpha\beta} \loarrow{\Xi}^{\mu} \Xi^\rho{}_\rho{}_{\mu} \loarrow{\Xi}_{\nu} \Xi^{\nu}
+\frac{1}{12}g_{\alpha\beta} \loarrow{\Xi}^{\mu} \Xi_{\mu}{}_\rho{}^\rho \loarrow{\Xi}_{\nu} \Xi^{\nu}
\nn&~~
-\frac{1}{12} \loarrow{\Xi}_{\beta\alpha}{}_{\mu} \Xi^{\mu} \loarrow{\Xi}_{\nu} \Xi^{\nu}
+\frac{1}{12} \loarrow{\Xi}_{\alpha}{}_{\mu\beta} \Xi^{\mu} \loarrow{\Xi}_{\nu} \Xi^{\nu}
\nn&~~
-\frac{1}{12} \loarrow{\Xi}_{\alpha} \Xi_{\beta\rho}{}^\rho \loarrow{\Xi}_{\nu} \Xi^{\nu}
+\frac{1}{12} \loarrow{\Xi}_{\alpha} \Xi_{\rho\beta}{}^\rho \loarrow{\Xi}_{\nu} \Xi^{\nu}
\nn&~~
-\frac{1}{12} \loarrow{\Xi}^{\mu} \Xi_{\beta\alpha\mu} \loarrow{\Xi}_{\nu} \Xi^{\nu}
+\frac{1}{12} \loarrow{\Xi}^{\mu} \Xi_{\alpha\mu\beta} \loarrow{\Xi}_{\nu} \Xi^{\nu}
\nn&~~
-\frac{1}{12} \loarrow{\Xi}_{\beta\rho}{}^\rho \Xi_{\alpha} \loarrow{\Xi}_{\nu} \Xi^{\nu}
+\frac{1}{12} \loarrow{\Xi}_{\rho\alpha}{}^\rho \Xi_{\beta} \loarrow{\Xi}_{\nu} \Xi^{\nu}
.
\end{align}
Upon using the commutator relations~(\ref{comcs1}-\ref{comcs2}) and the fact that the indices ${}_\alpha$ and ${}_\beta$ are symmetric, each line of the last 5 lines can separately be reduced to lower order derivative terms involving the gauge field strength or the curvature tensors. Using the same technique and neglecting all the lower order derivative terms, the first 3 lines and the second 3 lines can respectively be grouped together into the first and the second term of Eq.~(\ref{eomcs4}) but with the opposite sigh. Therefore all the higher order derivatives in $g^{\alpha\beta}$'s equation of motion exactly cancel. There is no Ostrogradski ghost for the modified Lagrangian (\ref{frgalcs}).

Now, we want to compare the differences that arise between the gauge and gravitational covariantisations. When applied to the galileon model in both cases, the partial derivatives are replaced by the corresponding covariant derivatives, which, unlike the partial derivatives, do not commute with themselves. Commuting two covariant derivatives gives rise to a term involving either the gauge field strength (gauge case) or the Riemann tensor (gravity case). In the covariantized theories, the inherited galileon anti-symmetrisation of the spacetime indices leads to a significant reduction in the number of higher order derivatives in the equations of motion, and indeed eliminates all the higher order derivatives for the gauge case. However, This total elimination does not occur in a gravitational covariantisation. The reason for the difference is due to the fact that, while the gauge field strength only contains first order derivatives, the Riemann tensor contains second order derivatives.

\section{Non-minimal gauging and general representations} \label{ggalext}

Now, let us discuss other ways to gauge the symmetric galileon with fundamental symmetries. We mentioned earlier that the Lagrangian (\ref{frgbsym}) is the only possible Lagrangian for the galileon globally charged by the fundamental representation of the classical groups. This is due to the fact that the partial derivatives commute with each other and the partial derivatives' spacetime indices are ``dressed'' with the anti-symmetrisation. Therefore the partial derivatives can be freely moved from one $\pi_{i_k}$ to another via integration by parts. However, this feature is not kept intact in the gauge covariantized theories, both in flat and curved space. In these covariantized theories, we can still do covariant integration by parts and the anti-symmetrisation ``dressing'' still largely reduces higher order derivatives, but the covariant derivatives do not commute and their commutators give rise to terms with the galileon coupled to the gauge field strength or the curvature tensors.

Indeed, in the flat space case, the following terms
\begin{align}
\label{nmc1}
&\left((D^\mu \pi)^\dagger D_{[\mu}D^\nu \pi \right) \left( (D_{\nu} \pi)^\dagger D^{\rho} D_{\rho]} \pi \right) + h.c.   ,
\\
\label{nmc2}
&\left((D^\mu \pi)^\dagger D_{[\mu}D^\nu \pi \right) \left( (D_{\nu} D^{\rho} \pi)^\dagger  D_{\rho]} \pi \right) ,
\end{align}
where $h.c.$ means the Hermitian conjugate of the previous terms, are different from the $\lambda_4$ term of (\ref{frgal}). Nevertheless, thanks to the anti-symmetrisation ``dressing'', these terms do not give rise to higher order derivative terms in the equations of motion, as one may explicitly check. Indeed, by integration by parts, these terms can be reduced to the $\lambda_4$ term of (\ref{frgal}) plus a number of terms involving non-minimal couplings to the field strength $F_{\mu\nu}$. The situation for the curved space case is similar. We will have the curved space counterpart of the terms (\ref{nmc1}) and (\ref{nmc2}) (with $D_\mu$ replaced by $\mathcal{D}_\mu$). Apart from the non-minimal couplings to the gauge field strength, non-minimal couplings to the curvature tensors will also arise from reducing the curved space counterpart of the terms (\ref{nmc1}) and (\ref{nmc2}) to the $\gamma_4$ term of (\ref{frgalcs}).

For representations other than the fundamental of the classical groups $SU(N)$, $SO(N)$ and $Sp(N)$, as we mentioned, more complicated galileon terms, such as those of (\ref{eq:sunAdjoint}), are allowed in the Lagrangian with global internal symmetry. When constructing the gauged theory, we will also have the choice of minimal gauging or non-minimal gauging. Nevertheless, in flat space, the gauged galileon theory, minimally or non-minimally gauged, do not have Ostrogradski ghosts in any representation of any gauge group, including the exceptional Lie groups. As one may have realized from our explicit calculations with the fundamental representation, the reason for this is that: In the equations of motion of both the galileon and the gauge field, the gauge covariant derivatives are ``dressed'' by the anti-symmetrisation of the spacetime indices and the most apparent higher order derivatives are fourth order; After using the commutator relations of the gauge covariant derivatives, there are only first order derivatives of the gauge field strength, which itself only has first order derivatives; Therefore in the equations of motion the most one can get is second order derivatives.

\section{Summary and Outlook}  \label{summout}

Although originally proposed as modifications to General Relativity on very large scales, the galileon models have taken on their own lives as effective field theories, due to their interesting field-theoretic properties. Various generalisations of the original galileon model have been proposed recently. In particular, galileon models with global internal symmetry have been presented and explicitly investigated in a probe brane setup with bulk isometries. In this paper, we have constructed for the first time the galileon models where the internal symmetry becomes local, i.e., the galileon is coupled to a Yang-Mills gauge field. This gauge covariantisation will break the Galilean field symmetry, akin to the gravitational covariantisation of~\cite{Deffayet:2009wt}. But in the limit where the local symmetry reduces to a global one, the galileon model with a global symmetry will be recovered. We have shown that this generalisation is natural in the sense that the Ostrogradski ghosts remain absent after the gauge covariantisation, despite apparent higher order derivatives in the Lagrangian. Note that it is indeed non-trivial for an apparent higher derivative Lagrangian term to be free of Ostrogradski ghosts. Although we only explicitly focused on the minimal gauging of the classical Lie groups' fundamental representation, we have shown that the absence of Ostrogradski ghosts is generally true for a gauged galileon theory in flat space in any representation of the gauge group. This is essentially because the gauge field strength only contains first order derivatives, while the Riemann tensor contains second order derivatives. We also couple the gauged galileon to gravity. For the naive covariantization, higher order derivatives do arise in the equations of motion. However, by adding an extra non-minimal coupling involving the Einstein tensor, the revised Lagrangian has a well-defined Cauchy problem, analogous to that of the covariant galileon~\cite{Deffayet:2009wt}.

We have argued that the gauged galileon can arise from a braneworld scenario, particularly from a probe brane setup. Galileon models with global internal symmetry arise when the bulk has maximal isometries, such as a Minkowski bulk with the induced brane metric $h_{\mu\nu} = \pd X^A/\pd x^\mu \pd X^B/\pd x^\nu G_{AB} = \eta_{\mu\nu}+\pd_\mu \pi^i \pd_\nu \pi_i$, where $X^A(x)$ is the embedding function and $G_{AB}$ is the bulk metric~\cite{Hinterbichler:2010xn}. However, generally, the bulk may have some unbroken general covariance, as well as possible isometries. In this case, the brane induced metric may obtain a form like $h_{\mu\nu} = g_{\mu\nu}+ (D_\mu \pi^i)^\dagger D_\nu \pi_i$, which leads to the gauged galileon in the fundamental representation we have explicitly considered. Simple counting of the degrees of freedom indicates that this kind of ansatz is generally obtainable, as there are more bulk metric components than the embedding equations needed to be satisfied, even for the Abelian case. It is interesting to identify the underlying symmetries of this braneworld setup and construct the corresponding bulk metric, which we leave for future work (Recently, Goon {\it et al} have explicitely realised our conjecture~\cite{Goon:2012mu}.).

We know gauge fields play a vital role in many self-consistent field theories and generally arise in stingy scenarios with extra dimensions, so galileon models must also be able to accommodate internal gauge symmetries. The goal of this paper is to demonstrate how this accommodation can be successfully realized. In~\cite{Padilla:2010ir}, we explicitly constructed a solitonic solution in the global SO(4) galileon model. The obtained gauged galileon terms may add extra features in constructing interesting non-perturbative field theory solutions such as monopoles, textures and cosmic strings.  The presence of such Abelian and non-Abelian gauge fields can also have dramatic consequences in the associated cosmology, ranging from the generation of primordial or cosmic scale magnetic fields to the formation and evolution of cosmic strings. We may also consider the gauged galileon models in curved space as modified gravity models, in which case the gauge field provide an extra vector mode, so we will have a scalar-vector-tensor theory. The consistencies of these models in different representations of the gauge group are yet to be investigated.

~\\

{\bf Acknowledgements}:~We would like to thank Paul Saffin, Andrew Tolley, Claudia de Rham, Antonio Padilla, Ian Kimpton and Zong-Gang Mou for useful discussions. EJC acknowledges support from the Royal Society and the Leverhulme Trust. SYZ thanks support from the University of Nottingham.


\begin{thebibliography}{99}

\bibitem{Trodden:2011xh}
  M.~Trodden, K.~Hinterbichler,
  Class.\ Quant.\ Grav.\  {\bf 28 } (2011)  204003;
  1104.2088 [hep-th].


\bibitem{Clifton:2011jh}
  T.~Clifton, P.~G.~Ferreira, A.~Padilla, C.~Skordis,
%
  1106.2476 [astro-ph.CO].
  

\bibitem{Nicolis:2008in}
  A.~Nicolis, R.~Rattazzi, E.~Trincherini,
  Phys.\ Rev.\  {\bf D79 } (2009)  064036;
  0811.2197 [hep-th].  


\bibitem{Dvali:2000hr}
  G.~R.~Dvali, G.~Gabadadze, M.~Porrati,
  Phys.\ Lett.\  {\bf B485 } (2000)  208-214;
  hep-th/0005016.


\bibitem{Luty:2003vm}
  M.~A.~Luty, M.~Porrati, R.~Rattazzi,
  JHEP {\bf 0309 } (2003)  029;
  hep-th/0303116.

\bibitem{Nicolis:2004qq}
  A.~Nicolis, R.~Rattazzi,
  JHEP {\bf 0406 } (2004)  059;
  hep-th/0404159.


\bibitem{ostroghost}
  R.~P.~Woodard,
  Lect.\ Notes Phys.\  {\bf 720 } (2007)  403-433;
  astro-ph/0601672.
  C.~Deffayet, J.~-W.~Rombouts,
  Phys.\ Rev.\  {\bf D72 } (2005)  044003;
  gr-qc/0505134.

\bibitem{dgpghost}
  C.~Charmousis, R.~Gregory, N.~Kaloper, A.~Padilla,
  JHEP {\bf 0610 } (2006)  066;
  hep-th/0604086.
  K.~Koyama,
  Phys.\ Rev.\  {\bf D72 } (2005)  123511;
  hep-th/0503191 [hep-th].

\bibitem{Hinterbichler:2010xn}
  K.~Hinterbichler, M.~Trodden, D.~Wesley,
  Phys.\ Rev.\  {\bf D82 } (2010)  124018;
  1008.1305 [hep-th].

\bibitem{Boulware:1973my}
  D.~G.~Boulware and S.~Deser,
  Phys.\ Rev.\ D {\bf 6} (1972) 3368.


\bibitem{arXiv:1006.4367} 
  C.~de Rham and G.~Gabadadze,
  Phys.\ Lett.\ B\ {\bf 693}  (2010) 334;
  1006.4367 [hep-th].

\bibitem{arXiv:1007.0443} 
  C.~de Rham and G.~Gabadadze,
  Phys.\ Rev.\ D\ {\bf 82}  (2010) 044020;
  1007.0443 [hep-th].

\bibitem{deRham:2010kj}
  C.~de Rham, G.~Gabadadze and A.~J.~Tolley,
  Phys.\ Rev.\ Lett.\  {\bf 106} (2011) 231101;
  1011.1232 [hep-th].


\bibitem{HassanRosen}
  S.~F.~Hassan and R.~A.~Rosen,
  Phys.\ Rev.\ Lett.\  {\bf 108}, 041101 (2012);
  1106.3344 [hep-th].
  S.~F.~Hassan and R.~A.~Rosen,
  1111.2070 [hep-th].


\bibitem{psz12}
  A.~Padilla, P.~M.~Saffin, S.~-Y.~Zhou,
  JHEP {\bf 1012 } (2010)  031;
  1007.5424 [hep-th].
  A.~Padilla, P.~M.~Saffin, S.~-Y.~Zhou,
  JHEP {\bf 1101 } (2011)  099;
  1008.3312 [hep-th].


\bibitem{Deffayet:2010zh}
  C.~Deffayet, S.~Deser, G.~Esposito-Farese,
  Phys.\ Rev.\  {\bf D82 } (2010)  061501;
  1007.5278 [gr-qc].

\bibitem{Padilla:2010ir}
  A.~Padilla, P.~M.~Saffin, S.~-Y.~Zhou,
  Phys.\ Rev.\  {\bf D83 } (2011)  045009;
  1008.0745 [hep-th].

\bibitem{Zhou:2010di}
  S.~-Y.~Zhou,
  Phys.\ Rev.\  {\bf D83 } (2011)  064005;
  1011.0863 [hep-th].

\bibitem{Andrews:2010km}
  M.~Andrews, K.~Hinterbichler, J.~Khoury, M.~Trodden,
  Phys.\ Rev.\  {\bf D83 } (2011)  044042;
  1008.4128 [hep-th].

\bibitem{Khoury:2011da}
  J.~Khoury, J.~-L.~Lehners, B.~A.~Ovrut,
  Phys.\ Rev.\  {\bf D84 } (2011)  043521;
  1103.0003 [hep-th].

\bibitem{Deffayet:2011gz}
  C.~Deffayet, X.~Gao, D.~A.~Steer, G.~Zahariade,
  Phys.\ Rev.\  {\bf D84 } (2011)  064039;
  1103.3260 [hep-th].

\bibitem{Charmousis:2011bf}
  C.~Charmousis, E.~J.~Copeland, A.~Padilla, P.~M.~Saffin,
%
  1106.2000 [hep-th].


\bibitem{horndeski}
G.~W.~Horndeski,
Int.\ J.\ Theor.\ Phys.\ {\bf 10} (1974) 363-384.


\bibitem{Deffayet:2009wt}
  C.~Deffayet, G.~Esposito-Farese, A.~Vikman,
  Phys.\ Rev.\  {\bf D79 } (2009)  084003;
  0901.1314 [hep-th].


\bibitem{deRham:2010eu}
  C.~de Rham, A.~J.~Tolley,
  JCAP {\bf 1005 } (2010)  015;
  1003.5917 [hep-th].

\bibitem{GHT2011}
  G.~Goon, K.~Hinterbichler, M.~Trodden,
  Phys.\ Rev.\ Lett.\  {\bf 106 } (2011)  231102;
  1103.6029 [hep-th].
  G.~Goon, K.~Hinterbichler, M.~Trodden,
  JCAP {\bf 1107 } (2011)  017;
  1103.5745 [hep-th].
  C.~Burrage, C.~de Rham and L.~Heisenberg,
  JCAP\ {\bf 1105}, 025  (2011);
  1104.0155 [hep-th].

\bibitem{Endlich:2010zj}
  S.~Endlich, K.~Hinterbichler, L.~Hui, A.~Nicolis, J.~Wang,
  JHEP {\bf 1105 } (2011)  073;
  1002.4873 [hep-th].


\bibitem{Babichev:2010kj}
  E.~Babichev,
  Phys.\ Rev.\  {\bf D83 } (2011)  024008;
  1009.2921 [hep-th].

\bibitem{Hui:2010dn}
  L.~Hui, A.~Nicolis,
  Phys.\ Rev.\ Lett.\  {\bf 105 } (2010)  231101;
  1009.2520 [hep-th].

\bibitem{Nicolis:2010se}
  A.~Nicolis,
%
  1011.3057 [hep-th].

\bibitem{Pujolas:2011he}
  O.~Pujolas, I.~Sawicki and A.~Vikman,
  JHEP {\bf 1111} (2011) 156;
  1103.5360 [hep-th].


\bibitem{latecos}
  N.~Chow, J.~Khoury,
  Phys.\ Rev.\  {\bf D80 } (2009)  024037;
  0905.1325 [hep-th].
  F.~PSilva, K.~Koyama,
  Phys.\ Rev.\  {\bf D80 } (2009)  121301;
  0909.4538 [astro-ph.CO].
  T.~Kobayashi, H.~Tashiro and D.~Suzuki,
  Phys.\ Rev.\ D {\bf 81} (2010) 063513;
  0912.4641 [astro-ph.CO].
  T.~Kobayashi,
  Phys.\ Rev.\ D {\bf 81} (2010) 103533;
  1003.3281 [astro-ph.CO].
  R.~Gannouji, M.~Sami,
  Phys.\ Rev.\  {\bf D82 } (2010)  024011;
  1004.2808 [gr-qc].
  A.~De Felice, R.~Kase, S.~Tsujikawa,
  Phys.\ Rev.\  {\bf D83 } (2011)  043515;
  1011.6132 [astro-ph.CO].
  C.~de Rham, L.~Heisenberg,
  Phys.\ Rev.\  {\bf D84 } (2011)  043503;
  1106.3312 [hep-th].
  C.~Deffayet, O.~Pujolas, I.~Sawicki and A.~Vikman,
  JCAP {\bf 1010} (2010) 026;
  1008.0048 [hep-th].



\bibitem{earlycos}
  P.~Creminelli, A.~Nicolis, E.~Trincherini,
  JCAP {\bf 1011 } (2010)  021;
  1007.0027 [hep-th].
  T.~Kobayashi, M.~Yamaguchi and J.~'i.~Yokoyama,
  Phys.\ Rev.\ Lett.\  {\bf 105} (2010) 231302;
  1008.0603 [hep-th].
  S.~Mizuno, K.~Koyama,
  Phys.\ Rev.\  {\bf D82 } (2010)  103518;
  1009.0677 [hep-th].
  C.~Burrage, C.~de Rham, D.~Seery, A.~J.~Tolley,
  JCAP {\bf 1101 } (2011)  014;
  1009.2497 [hep-th].
  Z.~-G.~Liu, J.~Zhang, Y.~-S.~Piao,
  Phys.\ Rev.\  {\bf D84 } (2011)  063508;
  1105.5713 [astro-ph.CO].
  T.~Qiu, J.~Evslin, Y.~-F.~Cai, M.~Li, X.~Zhang,
  JCAP {\bf 1110 } (2011)  036;
  1108.0593 [hep-th].
  X.~Gao, D.~A.~Steer,
%
  1107.2642 [astro-ph.CO].
  L.~Levasseur Perreault, R.~Brandenberger, A.~-C.~Davis,
%
  1105.5649 [astro-ph.CO].
  D.~A.~Easson, I.~Sawicki and A.~Vikman,
  JCAP {\bf 1111} (2011) 021;
  1109.1047 [hep-th].
  H.~Wang, T.~Qiu and Y.~-S.~Piao,
  Phys.\ Lett.\ B {\bf 707}, 11 (2012);
  1110.1795 [hep-ph].


\bibitem{arXiv:1103.1351} 
  C.~de Rham, G.~Gabadadze, D.~Pirtskhalava, A.~J.~Tolley and I.~Yavin,
  JHEP\ {\bf 1106}  (2011) 028;
  1103.1351 [hep-th].


\bibitem{Fuchs:1997jv}
  J.~Fuchs, C.~Schweigert,
  Cambridge, UK: Univ.\ Pr.\ (1997) 438 p.

\bibitem{weinberg}
S.~Weinberg, New York, USA: John Wiley \& Sons (1972) 657 p.

\bibitem{Goon:2012mu}
  G.~Goon, K.~Hinterbichler, A.~Joyce and M.~Trodden,
  arXiv:1201.0015 [hep-th].



\end{thebibliography}
\end{document}